\documentstyle{elsart}

\begin{document}
\title{A Note on "Optimal Static Load Balancing in Distributed Computer Systems"}

\begin{center}
Sakib A. Mondal\footnote[1]{Address for Correspondence: Infosys Technologies Limited,\\
\hspace*{0.22in}27, Bannerghatta Road, JPNagar 3rd Phase,\\
\hspace*{0.22in}Bangalore 560 076,India Fax 91-080-6587967\\
\hspace*{0.25in}email: abdulsakib@inf.com, sakib@hotmail.com}\\
Infosys Technologies Limited
\end{center}

{\bf Abstract:} The problem of minimizing mean response time of generic jobs submitted to a heterogenous distributed computer systems is considered in this paper.  A static load balancing strategy, in which decision of redistribution of loads does not depend on the state of the system, is used for this purpose. The article is closely related to a previous article on the same topic. The present article points out number of inconsistencies in the previous article, provides a new formulation, and discusses the impact of new findings, based on the improved formulation, on the results of the previous article.

\section{Introduction}
Recent years have been witness to an increasing use of distributed computing system. This may be attributed to two main factors: growth of the Internet, and low cost solution of end-user computing devices. Many business processes such as supply chain management are distributed due to the inherent nature of tasks involved with them.
Besides, scale of economy is often possible due to the use of clusters of less powerful computers instead of a central computer of significantly high power. However, a distributed solution can yield the true advantage only if it is possible to distribute works evenly among the processors (terms processor, host and computer are used inter-changeably in this article). In other words, when load on the computers in a distributed environment has significant variance of workloads, high performance can be achieved by redistributing loads. The task of redistributing the loads on the computers is called load balancing.

Load can be characterized as jobs or tasks. A job indicates a complete and independent entity of work to be completed. However,  a job may consist of number of sub-units called tasks, and for the job to be completed the comprising tasks need to often communicate among themselves. For simplification, in this article, we consider mutually independent tasks units as jobs. Jobs can be classified into two broad categories - dedicated and generic. Dedicated jobs can be processed only on specified hosts, while generic jobs can be processed on any host in the system. Once a generic job is submitted to the system, it can be processed at the place of its origin or  can be transferred it to another processor. Irrespective of the processor chosen, once the job is started at a processor, it remains there till completion. Some of the most commonly used performance indicators of a load balancing algorithm are mean response time, throughput, variance of response time. In this article, we confine our analysis based on mean response time only.

Algorithms or heuristics for load balancing fall into two categories: static and dynamic. Static load balancing techniques do not depend on the state (characterized by workload etc.) of the processors. Simple static load balancing techniques such as join-the-shortest queue (SQ), minimum expected delay (MED) have been in use for over a decade. The SQ policy allocates an arriving task to the processor having the minimum load and, the MED policy allocates an arriving task to the processor having the minimum expected value of waiting time for currently scheduled tasks. Note that though these two policies may work similarly for homogenous processors, their behaviors differ for heterogeneous processors. Tantawi and Towsley \cite{ta85} have proposed an optimization model and load balancing algorithms to determine static optimal allocation of loads among the hosts connected in a broadcast network. Ross et al.  \cite{ro91}
considered a more general problem consisting of dedicated and generic jobs, and also dealt with scheduling decision at each host. The authors have noted that the problem is separable over local scheduling decisions, and suggested a solution procedure based on this finding.
  
Dynamic load balancing techniques, on the other hand, rely on present and past states of the processors. In dynamic load balancing, though an initial assignment of work to the  processors is done through static load balancing, subsequent
adjustment of work depends on the workload profile of the processors.
Dynamic load balancing is expected to have higher overhead, but better
performance than static load balancing. Most of the dynamic load balancing algorithms have two important components: Information exchange through which processors exchange their load information, and subsequent load exchange
to redistribute loads. Depending on the characteristics of these components,  there can be different types of dynamic load balancing ( \cite{de98} \cite{le97}, \cite{mo00},  \cite{za97}).

However, in this article we concentrate on static load balancing. We express some of the points of our disagreement with \cite{ta85} and
provide alternatives. The rest of the paper is organized as follows. Section 2 suggests an improved problem formulation. The solution procedure is illustrated in Section 3. Concluding remarks are provided in Section 4. Points of disagreement are marked in italics.

\section{ Improved Problem Formulation}
We adopt the model of distributed computing as proposed in \cite{ta85}. In this model host computers are treated as node. Each node $i \in \{1,2, \cdots n\}$ is capable of executing any incoming job k. Arrival rate of job at node i is $\phi_i$, and total arrival rate of jobs on the system is $\Phi = \sum_{i=1}^n \phi_i$. When a job k arrive at a node i, it may be executed at node i itself or transferred to another node j, depending on the distribution of loads at that time. If job is transferred to node j, it is executed at node j without being transferred to any other host, and results of execution is transferred back to node i so that the user submitting the job gets the illusion of the job being executed at the host i. Processing rate of job at node i is $\beta_i$ and transfer rate of jobs from node i to node j is $x_{ij}$.  Therefore,
$\lambda =\sum_{i=1}^{n} \sum_{j=1}^{n} x_{ij}$ is the total traffic on the network.

It is also assumed that hosts may be heterogeneous (different CPU speed, memory and swap availability etc.) and speed of execution depends on the instantaneous load of the host. Therefore the instantaneous marginal delay due to local processing of a job k at host i depends on host i as well as load on the host i, and is denoted by $F_i(\beta i)$.
So the mean response time of all the jobs on the network due to local processing is given by
$\sum_{i=1}^n \frac{\beta_i}{\sum_{i=1}^{n}\beta_i} F_i(\beta i) =
\sum_{i=1}^n \frac{\beta_i}{\beta} F_i(\beta i)$.

In the distributed set up, another major component of mean response time is the network delay. The incremental communication delay for transfer of a packet can be modeled as $G_{ij}({\bf x})$, where ${\bf x}$ is the vector of load transfer. Note that we are assuming that the network delay due to sending the response back is negligible. In case it is not so, the argument of G should be suitably replaced. \footnote{If we assume that $y_{ij}$ is the
amount of flow of results from node i to node j, then a more accurate model would have network delay given by:
$\sum_{i=1}^n \sum_{j=1}^{n} \frac{x_{ij}+y_{ji}}{\sum_{i=1}^{n} \sum_{j=1}^{n} (x_{ij}+y_{ji})} G_{ij}({\bf x}+{\bf y})$, where
${\bf y}$ is the vector of additional traffic due to response packets.}
So the contribution of network delay to mean response time is given by
$\sum_{i=1}^n \sum_{j=1}^{n} \frac{x_{ij}}{\sum_{i=1}^{n} \sum_{j=1}^{n} x_{ij}} G_{ij}({\bf x})$. 

[{\em It is to be noted that in \cite{ta85} as well as in \cite{ro91}, denominator used was $\Phi $ which is not equal to $\sum_{i=1}^n \sum_{j=1}^n x_{ij}$ . Hence, their formulations did not use average communication delay, contrary to their claims. Also the claim in \cite{ta85} that ${\bf x}$ in $G({\bf x})$  can include response packets is wrong as in that case total flow no longer remains $\lambda$.}]

Hence the problem of load balancing can be formulated as:

Minimize 
\[\sum_{i=1}^n \frac{\beta_i}{\sum_{i=1}^{n} \beta_i} F_i(\beta i)+
\sum_{i=1}^n \sum_{j=1}^{n} \frac{x_{ij}}{\sum_{i=1}^{n} \sum_{j=1}^{n} x_{ij}} G_{ij}({\bf x})\]
subject to
\[ \sum_{i=1}^{n} \beta_i = \sum_{i=1}^{n}\phi_i ,\]
\[ \beta_i + \sum_{j=1}^{n} x_{ij} = \phi_i + \sum_{j=1}^n x_{ji} , \; i = 1, 2, \cdots , n \]
\[ \lambda =\sum_{i=1}^{n} \sum_{j=1}^{n} x_{ij},\]
\[ \beta_i \geq 0 , \; i = 1, 2, \cdots , n,\]
\[ x_{ij} \geq 0 , \; i,j = 1, 2, \cdots , n,\]
\[ x_{ii} = 0 , \; i = 1, 2, \cdots , n,\]
\begin{equation}
\label{P1}
\end{equation}

Note that first two constraints ensure that all incoming load is processed, and flow balance holds true.  As in \cite{ta85}, we assume that $F_i(\beta_i)$ are increasing convex functions  and $G_{ij}({\bf x})$ is a non-decreasing convex function. Ethernet and satellite are quite common network infrastructure. Due to the type of data transmission means on these, the network delay for transfer of a packet from node i to node j depends on the total network load. Henceforth, we also assume that $G_{ij}({\bf x})$ depends on the total
traffic $\lambda $ on the network but not on any component of it. So the second term in the objective function can be written as $ \sum_{i=1}^n \sum_{j=1}^{n} \frac{x_{ij}}{\lambda} G_{ij}(\lambda)$. 

\section{Solution of the Problem}
Solution to the above problem can be expressed in terms of disjoint set of nodes. We classify the nodes as source (idle or active), neutral and sinks as in \cite{ta85}. A node $i$ is said to be source if $x_{ji}=0 \; \forall j$ and $\exists j \; x_{ij}>0$. A source $i$ is  idle if $\beta_i=0$, else it is active. A node $i$ is said to be sink if $x_{ij}=0 \; \forall j$ and $\exists j \; x_{ji}>0$. For a neutral node $i$,  $x_{ij}=0 \; \forall j$ and $x_{ji}=0 \; \forall j$. We denote the set of sinks, idle sources, active sources and neutrals by $S, R_d, R_a$ and $N$ respectively.

Tantawi and Towsley \cite{ta85} claimed that triangular inequality of network delays (i.e., $G_{ij}(\lambda ) \leq G_{ik} (\lambda ) +G_{kj} (\lambda ) $) was sufficient condition for solution of the problem as formulated by them to yield only source, sink and neutral nodes. We illustrate in the following theorem that the claim is wrong.

\begin{thm} The triangular inequality of network delay and non-decreasing  property of $\frac{G(\lambda)}{\lambda}$ are sufficient conditions for Problem~\ref{P1} to yield nodes which are either sources, sinks or neutrals.
\end{thm}

{\bf Proof:} For a given incoming load $\beta$ is constant, and hence we just need to show that the stated conditions are sufficient for a flow rate assignment ${\bf x}$ with nodes either sinks, sources or neutral nodes to minimize the communication term $C= \sum_{i=1}^n \sum_{j=1}^{n} \frac{x_{ij}}{\lambda} G_{ij}(\lambda)$. Let us suppose that $i({\bf x})$ denotes the number of nodes that are neither sinks, sources nor neutrals for the flow rate assignment ${\bf x}$. Therefore we need to show that  $i({\bf x})=0$. Of all the flow rate assignment minimizing communication delay, choose the flow rate assignment ${\bf x}$  which has maximum $i({\bf x})$ and it is strictly positive. [{\em In Proof of Theorem 1 in \cite{ta85}, ${\bf x}$ was chosen to have minimum $i({\bf x})$ from candidate solutions minimizing communication delay. And it was proved that $i({\bf x})=0$. But this does not rule out the possibility that there can be other optimal solutions with $i({\bf x})>0$. So it only proves that there is at least one flow rate assignment minimizing the objective function for which nodes are either sources, sinks or neutrals. }] Since $i({\bf x})>0$, there is at least one node $k$ and a pair $l$ and $m$ such that $x_{lk}>0$ and $ x_{km}>0$. Consider a new assignment ${\bf x'}$ which differs from ${\bf x}$  only in the following. 
\[x_{lk}'= x_{lk}-\min \{x_{lk}, x_{km}\},\] 
 \[x_{km}'= x_{km}-\min \{x_{lk}, x_{km}\},\] 
 \[x_{lm}'= x_{lm}+\min \{x_{lk}, x_{km}\}.\] 
Let the new Communication delay be $C'$. So,
\[C'=\sum_{i=1}^n \sum_{j=1}^{n} \frac{x_{ij}'}{\lambda '} G_{ij}(\lambda ')=\]  \[ x_{lk}' \frac{ G_{lk}(\lambda ')}{\lambda '}+x_{km}' \frac{ G_{km}(\lambda ')}{\lambda '}+x_{lm}' \frac{ G_{lm}(\lambda ')}{\lambda '}+ (\sum_{i=1}^n \sum_{j=1}^{n} x_{ij} \frac{ G_{ij}(\lambda ')}{\lambda '} \mid (i,j) \not \in (l,k), (k,m), (l,m) \} ).\]
Since $\lambda ' \leq \lambda$ and $\frac{G(\lambda)}{\lambda}$ is non-decreasing, 
\[C'-C \leq \{ x_{lk}' \frac{ G_{lk}(\lambda ')}{\lambda '} - x_{lk} \frac{ G_{lk}(\lambda)}{\lambda} \} + \{ x_{km}' \frac{ G_{km}(\lambda ')}{\lambda '} - x_{km} \frac{ G_{km}(\lambda)}{\lambda }\}\]
\[+ \{ x_{lm}' \frac{ G_{lm}(\lambda ')}{\lambda '} -  x_{lm} \frac{ G_{lm}(\lambda)}{\lambda }\}\leq  ( x_{lk}' -x_{lk})  \frac{ G_{lk}(\lambda)}{\lambda} \]
\[+ ( x_{km}' - x_{km}) \frac{ G_{km}(\lambda)}{\lambda } 
+ ( x_{lm}'-x_{lm}) \frac{ G_{lm}(\lambda)}{\lambda }\]
\[= \frac{-1}{\lambda} [\min \{x_{lk}, x_{km}\} (G_{lk}(\lambda ) + G_{km} (\lambda ) -G_{lm} (\lambda ))] \leq 0.\]
Therefore $i({\bf x}')<i({\bf x})$, a contradiction. 

[{\em in Proof of Theorem 1 in \cite{ta85}, with new ${\bf x '}$ total network load no longer remains $\lambda$. Also it is possible to have $i({\bf x}')=i({\bf x})-2$, and not always $i({\bf x}')=i({\bf x})-1)$.}]
\qed

As for broadcast network, delay can be assumed to be independent of source-destination pair, $G_{ij} (\lambda)$ is independent of i and j. Moreover, since $\sum_{i=1}^{n} \beta_i =\sum_{i=1}^{n} \phi_i$ and total incoming load is constant, the objective function in Problem~\ref{P1} is equivalent to \[ \sum_{i=1}^n \beta_i F_i (\beta_i) + \sum_{i=1}^n \beta_i G(\lambda ).\]

For simplifying the solution procedure, let $u_i =\sum_{j=1}^n x_{ji},
v_i =\sum_{j=1}^n x_{ij}$. So Problem~\ref{P1} can be written as:

Minimize 
\[ \sum_{i=1}^n \beta_i F_i (\beta_i) + \sum_{i=1}^n \beta_i G(\sum_{i=1}^n v_i)\]
subject to
\[ \beta_i + v_i = \phi_i +u_i , \; i = 1, 2, \cdots , n,\]
\[ \sum_{i=1}^{n} v_{i} -\sum_{i=1}^n u_{i}=0 ,  \; i = 1, 2, \cdots , n \]
\[ \beta_i \geq 0 , \; i = 1, 2, \cdots , n,\]
\[ u_{i} \geq 0 , \; i = 1, 2, \cdots , n,\]
\[ v_{i} \geq 0 , \; i = 1, 2, \cdots , n.\]
\begin{equation}
\label{P2}
\end{equation}

Note that $\sum_{i=1}^n \beta_i=\sum_{i=1}^n \phi_i=\Phi$ is constant. After eliminating the first constraint, problem~\ref{P2} has the following form.

Minimize 
\[ \sum_{i=1}^n (u_i-v_i+\phi_i) F_i (u_i-v_i+\phi_i) + \Phi G(\sum_{i=1}^n v_i)\]
subject to
\[ \sum_{i=1}^{n} v_{i} -\sum_{i=1}^n u_{i}=0 , \; i = 1, 2, \cdots , n \]
\[u_i - v_i + \phi_i \geq 0 ,\;  i = 1, 2, \cdots , n,\]
\[ u_{i} \geq 0 ,\; i = 1, 2, \cdots , n,\]
\[ v_{i} \geq 0 , \; i = 1, 2, \cdots , n.\]
\begin{equation}
\label{P3}
\end{equation}

\begin{thm}

The optimal solution to the above problem can be expressed as:
\[f_i(\beta_i) \geq \alpha+\beta G'(\lambda), \; \beta_i=0 \; (i\in R_d), \] 
\[ f_i(\beta_i) = \alpha + \beta G'(\lambda), \; 0<\beta_i <\phi_i \;  (i\in R_a), \]
\[\alpha \leq f_i(\beta_i) \leq \alpha + \beta G'(\lambda), \; \beta_i=\phi_i \; (i\in N), \]
\[f_i(\beta_i) =\alpha, \; \beta_i >\phi_i \; (i\in S), \]
subject to total flow constraint,
\[\sum_{i\in R_a}f_{i}^{-1}(\alpha + \Phi G'(\lambda))+ \sum_{i\in N} \phi_i + \sum_{i\in R_a}f_{i}^{-1}(\alpha) = \Phi,\] where 
\[\lambda= \sum_{i \in S} (f_{i}^{-1}(\alpha) - \phi_i).\]
\end{thm}

{\bf Proof:}
The Lagrangean function for this problem is
\[ L({\bf u},{\bf v}, \alpha, {\bf \gamma}, {\bf \psi},  {\bf \eta})=\sum_{i=1}^n (u_i-v_i+\phi_i) F_i (u_i-v_i+\phi_i) + \Phi G(\sum_{i=1}^n v_i)\]
\[+ \alpha ( \sum_{i=1}^{n} v_{i} -\sum_{i=1}^n u_{i}) + \sum_{i=1}^n \gamma_i (u_i-v_i+\phi_i) +\sum_{i=1}^n \psi_i u_i +\sum_{i=1}^n \eta_i v_i .\]

It can be shown easily that objective function and the constraints satisfy conditions for applying Karush-Kuhn-Tucker (KKT) conditions. Let $f_i(\beta_i)=F_i(\beta_i)+\beta_i \frac{ F_i(\beta_i)}{ \beta_i }$. Applying KKT conditions, we get the following set of equations.

\begin{equation}
 \frac{\delta L}{\delta u_i}=f_i(u_i - v_i + \phi_i) -\alpha + \gamma_i + \psi_i =0 ,\; i = 1, 2, \cdots , n, \label{L1}
\end{equation}

\begin{equation} \frac{\delta L}{\delta v_i}=-f_i(u_i - v_i + \phi_i) +\Phi 
G'(\sum_{i=1}^n v_i) 
+ \alpha - \gamma_i + \eta_i =0 , i = 1, 2, \cdots , n, \label{L2}
\end{equation}

\begin{equation}  \frac{\delta L}{\delta \alpha}= -\sum_{i=1}^n u_i + \sum_{i=1}^n v_i = 0 , \label{L3}
\end{equation}
\begin{equation}  u_i-v_i+\phi_i \geq 0,  \gamma_i(u_i-v_i+\phi_i)=0, \gamma_i \leq 0 ,\; i = 1, 2, \cdots , n, \label{L4}
\end{equation}
\begin{equation}  
u_{i} \geq 0 , \psi_i u_i =0, \psi_i \leq 0, \; i = 1, 2, \cdots , n, \label{L5}
\end{equation} 
\begin{equation}   v_{i} \geq 0 , \eta_i v_i = 0, \eta_i \leq 0,\; i = 1, 2, \cdots , n. \label{L6}
\end{equation} 

Let us consider two cases separately, $u_i-v_i+\phi_i =0$ and $u_i-v_i+\phi_i>0$. 
Case 1: $u_i-v_i+\phi_i =0$, i.e., $\beta_i=0$.

Again we consider two sub cases:

1A: $\phi_i >0$.

In this case $v_i>0$, and consequently  from Equation~\ref{L6} $\eta_i=0$.So from Equation~\ref{L2}
\[f_i(\beta_i) \geq \alpha+\beta G'(\lambda).\] Also for this case, $\beta_i=0, \phi_i>0$.
It can easily be noticed that these are idle source nodes.

1B: $\phi_i =0$.

It is obvious that in this case $u_i=v_i$. From Equation~\ref{L1} and Equation~\ref{L2},

\[ \Phi G'(\lambda)+ \gamma_i + \eta_i =0 .\]

But by assumption $\Phi G'(\lambda) > 0 $ and from Equation~\ref{L4} and Equation~\ref{L6} $\gamma_i \leq 0, \eta_i \leq 0$. This implies atleast one of $\gamma_i$  and $\eta_i$ is strictly negative, and correspondingly from Equation~\ref{L4} or Equation~\ref{L6} either  $u_i$ or $v_i$ is zero. Therefore $u_i= v_i =0$. Hence, from Equation~\ref{L2} $f_i(\beta_i) \geq \alpha.$ Also $\beta_i=0, \phi_i=0$.

These nodes correspond to neutral nodes without external load.

Case 2: $u_i-v_i+\phi_i >0$.

From Equation~\ref{L4}, $\gamma_i=0$. Now consider three subcases:

2A: $v_i >0, u_i=0$.

Since $v_i>0$, from Equation~\ref{L6} $\eta_i=0$. Hence from Equation~\ref{L2}, we get

\begin{equation}    f_i(\beta_i) = \alpha + \beta G'(\lambda). \label{L9}
\end{equation}   
For this case, $0<\beta_i <\phi_i$. Therefore these nodes correspond to active sources.

2B: $v_i=0, u_i>0$. 

Since $u_i>0$, from Equation~\ref{L5} $\psi_i=0$. Hence from Equation~\ref{L1},
\[f_i(\beta_i) =\alpha .\]   In this case $\beta_i >\phi_i$. These are sink nodes. 

2C. $u_i=v_i=0$.

Since $\psi_i \leq 0$, from Equation~\ref{L1} we get
$f_i(\beta_i) \geq \alpha$.  Similarly, since $\eta_i \leq 0$, from Equation~\ref{L2} we get $f_i(\beta_i) \leq \alpha + \beta G'(\lambda)$. Here $\beta_i=\phi_i$, and hence nodes are neutrals.

To solve the system completely, we need to find out value of $\alpha$. From Equation~\ref{L3}, $\sum_{i=1}^{n} \beta_i =\Phi$. Substituting values of $\beta_i$ from the solution nodes in the expression we get $\sum_{i\in R_a}f_{i}^{-1}(\alpha + \Phi G'(\lambda))+ \sum_{i\in N} \phi_i + \sum_{i\in R_a}f_{i}^{-1}(\alpha) = \Phi$.  We still need value of $\lambda$ to solve the above equation to find out $\alpha$ for the incumbent solution.  Fortunately $\lambda=\sum_{i=1}^n u_i = \sum_{i \in S} u_i =\sum_{i \in S} (\beta_i +v_i - \phi_i) =\sum_{i \in S} (\beta_i - \phi_i) =\sum_{i \in S} (f_{i}^{-1}(\alpha) - \phi_i)$. Now we are at a position to find out $\alpha$ from the above equation.

[{\em There are some mistakes in the proof of corresponding theorem in \cite{ta85} which we presume to be typographical error.}]
\qed 

The result of the above theorem can be intuitively explained as illustrated below. All the sink nodes have the lowest and equal incremental node delay($\alpha$). Loads are shared among the sink nodes such that all have equal incremental node delay. For all the active source nodes the incremental node delay equals sum of incremental node delay of sink nodes and total incremental communication delay at the present load ($\beta$). If the incremental node delay at a source is greater than this value then it is profitable, from the perspective of the objective to reduce mean response time, to transfer all the load of the source to the sink nodes, and these nodes correspond to idle source nodes. Note that load distribution among the active source nodes are such that all of them have equal incremental node delay($\alpha+\beta G'(\lambda)$), and these nodes process portion of their incoming load ($\phi_i$) transferring the remaining loads to sinks.  However, if at a node the incremental node delay is less than the above value, then it is not profitable to send any portion of the load to sink node and thereby incur communication delay. All the incoming load at such a node is processed by itself, and it is a neutral node. The above solution is soothing to practitioners' eyes also.  The optimal solution suggests a distribution such that nodes with lower incremental node delay are utilized more extensively than others.

\section{Conclusion}

This article points out a number of inconsistencies in \cite{ta85}, and improves on them. One of the nice aspects of the results in the present article is that the parametric analysis as well as single-point load balancing algorithm as suggested in \cite{ta85} can be applied with suitable modification of expressions from the solution given in the previous theorem.  It is also perceivable that due to the simplicity of the expression, the present procedures can not be more difficult than those in \cite{85}.

Fortunately, the present finding does not affect the results of  \cite{ro91}.  As mentioned before, Ross et al.  had considered a general problem consisting of generic as well as dedicated jobs, and included the task of scheduling jobs at each host. The authors showed that problem was separable over scheduling decisions. They also showed that given an allocation of the jobs on the hosts, the task of scheduling can be solved as a polymatroid optimization problem.  The present results in this paper only changes the allocation of the jobs on the hosts and does not violate any of the assumptions made in \cite{ro91}.

As noted in \cite{ta85}, dynamic load balancing can yield better response time. Most of the dynamic load balancing tool uses a decision making unit (called load balancer) which periodically monitors load on hosts. Also hosts can decide to inform of any exigency when load on them exceed a high level threshold or goes below a low level threshold (both these thresholds can be dynamic). Load balancer then decides on new load distribution and executes the transfer.  The present result can be used for dynamic load balancing. The expected values of $\alpha$ and $\alpha+\beta G'(\lambda)$ can serve as lower and higher thresholds.  The load balancer can keep track of incremental node delay (a function of allocated load on the node. See \cite{mo00}  for some illustrative measures of load) of hosts as well as incremental communication delay. When it receives any receiver-initiated request for a load, generated by load on the host going below the lower threshold), it can invoke single-point algorithm to reallocate the load. Similarly, it can redistribute load when it receives any sender-initiated request, generated by load on the host going above the upper threshold.

\end{document}